\documentclass[journal,romanappendices]{IEEEtran}

\usepackage{graphicx}
\usepackage[font=footnotesize]{caption}
\usepackage[font=footnotesize]{subcaption}
\usepackage{cite}
\usepackage[cmex10]{amsmath}
\usepackage{amsfonts,amssymb,amsthm}
\usepackage{mathrsfs}
\usepackage{enumerate}
\usepackage{mathtools}
\usepackage{hyperref}

\newtheorem{definition}{Definition} 
\newtheorem{proposition}{Proposition}
\newtheorem{corollary}{Corollary}

\newtheorem{lemma}{Lemma}

\theoremstyle{remark}
\newtheorem{remark}{Remark}

\newcommand{\negsp} {
\mkern-1mu
}

\newcommand{\shortminus} {
\scalebox{0.75}[1.0]{\( - \)}
}

\begin{document}

\title{Decoding Delay Performance of Random Linear Network Coding for Broadcast}

\author{Ioannis~Chatzigeorgiou and Andrea~Tassi
\thanks{%
Copyright~\textcopyright~2015 IEEE. Personal use of this material is permitted. However, permission to use this material for any other purposes must be obtained from the IEEE by sending a request to \href{mailto:pubs-permissions@ieee.org}{pubs-permissions@ieee.org}. 

I. Chatzigeorgiou is with the School of Computing and Communications, InfoLab21, Lancaster University, Lancaster LA1~4WA, United Kingdom (\mbox{e-mail:} \href{mailto:i.chatzigeorgiou@lancaster.ac.uk}{i.chatzigeorgiou@lancaster.ac.uk}).%

A. Tassi is with the Department of Electrical and Electronic Engineering, University of Bristol, United Kingdom (e-mail: \href{mailto:a.tassi@bristol.ac.uk}{a.tassi@bristol.ac.uk}).%
}%
}

\maketitle

\begin{abstract}
Characterization of the delay profile of systems employing random linear network coding is important for the reliable provision of broadcast services. Previous studies focused on network coding over large finite fields or developed Markov chains to model the delay distribution but did not look at the effect of transmission deadlines on the delay. In this work, we consider generations of source packets that are encoded and transmitted over the erasure broadcast channel. The transmission of packets associated to a generation is taken to be \mbox{deadline-constrained}, that is, the transmitter drops a generation and proceeds to the next one when a predetermined deadline expires. \mbox{Closed-form} expressions for the average number of required packet transmissions per generation are obtained in terms of the generation size, the field size, the erasure probability and the deadline choice. An upper bound on the average decoding delay, which is tighter than previous bounds found in the literature, is also derived. Analysis shows that the proposed framework can be used to fine-tune the system parameters and ascertain that neither insufficient nor excessive amounts of packets are sent over the broadcast channel.
\end{abstract}

\begin{IEEEkeywords}
Network coding, rateless coding, systematic, non-systematic, broadcast, multicast, delay, probability analysis.
\end{IEEEkeywords}


\section{Introduction}
\label{sec:intro}

Since the inception of fountain coding for the broadcast channel \cite{Byers02} and network coding for connected network topologies \cite{Ahlswede00}, the fundamental idea of transmitting linear combinations of packets, either from a source node or intermediate network nodes, was extensively investigated. Research demonstrated that linear combinations do not need to conform to design rules or deterministic patterns; coding based on random linear combinations is a capacity-achieving scheme for multicast connections \cite{Lun08}. The so-called \textit{randomized network coding} \cite{Ho03}, which is also referred to as \textit{random linear network coding} \cite{Ho06}, offers clear advantages over conventional forwarding and routing techniques. Protocols that exploit its properties have been proposed, including Avalanche \cite{Gkantsidis05} for wireline networks and MORE \cite{Chachulski07} for wireless networks. Random linear network coding for vehicle-assisted wireless broadcast has also been considered. For example, data downloading via infrastructure-to-vehicle connections, and vehicle-to-vehicle data sharing when vehicles travel in the same direction has been explored in \cite{Zhu15}, while data dissemination between vehicles that move in opposite directions has been studied in \cite{Liu16}.

In random linear network coding for broadcast, a transmitter segments data into \textit{generations} of $K$ source packets each. For a given generation, the transmitter broadcasts coded packets, which are obtained by linearly combining the $K$ source packets of that generation over a finite field. A receiver needs to recover $K$ linearly independent coded packets in order to reconstruct the $K$ source packets of the generation using Gaussian elimination. The average decoding delay experienced by a receiver is the mean time required for the recovery of a generation. The average decoding delay imposed to the system is the expected number of time steps needed by all receivers to decode a generation \cite{Nistor2011}. The average decoding delay of the system can be computed by observing the decoding delay at each receiver, recording the longest  delay for every generation, and averaging over a long run of generations.

If the packet mixing operations are over a large finite field, the randomly generated coded packets are linearly independent with high probability. Therefore, a receiver is likely to recover the $K$ source packets if it collects exactly $K$ coded packets. Based on this assumption, the broadcast case can be modeled as multiple independent unicast cases. This simplification facilitates the calculation of the average decoding delay of the system because the joint probability of all receivers decoding the source packets can be expressed as the product of the marginal decoding probabilities of all receivers. Eryilmaz \textit{et al.} \cite{Eryilmaz2006} adopted this approach, obtained an upper bound on the average decoding delay at a receiver, derived expressions for the average decoding delay of the system and proposed various scheduling strategies when receivers send feedback to acknowledge recovery of the source packets. Lucani \textit{et al.} \cite{Lucani2009_ITA} considered the case where each receiver regularly reports to the transmitter the number of linearly independent coded packets that are still missing. The average decoding delay at a receiver was computed using a Markov chain model and was incorporated in an optimization strategy, which minimizes the number of coded packets that are broadcast before receivers are encouraged to send acknowledgements. Heide \textit{et al.} \cite{Heide2009} also treated the broadcast case as disjoint unicast cases and computed the average delay per transmitted packet for both \mbox{non-systematic} and systematic network coding.

The effect of the field size on the average decoding delay at a receiver was investigated by Lucani \textit{et al.}\mbox{\cite{Lucani2009_Globecom, Lucani2012}}. The authors derived an upper bound on the decoding delay, which was combined with that obtained in \cite{Eryilmaz2006}, and demonstrated that binary network coding exhibits a negligibly longer decoding delay than \mbox{non-binary} network coding for an increasing generation size. Nistor \textit{et al.} \cite{Nistor2011} recognized that the average decoding delay of a broadcast system can be easily computed only when specific channel conditions are met and discussed the complexity of deriving a general expression for the joint probability of all receivers decoding the source packets. To facilitate the analysis, the authors focused on a system comprising one transmitter and two receivers, proposed a Markov chain model to study the delay distribution of the system and showed that their model reduces to a Markov chain that is similar to that in \cite{Lucani2009_Globecom} when only one receiver is present.

In summary, the aforementioned literature on the delay performance of network-coded transmission over the broadcast channel either considered operations over large finite fields to simplify the analysis or resorted to Markov chains to model the delay distribution. The underlying hypothesis that previous studies have in common is that a receiver always collects the required number of linearly independent coded packets and recovers a generation of source packets. In this paper, we consider a transmitter that abides by a deadline, after which coded packets related to a generation are no longer broadcast. In particular, the contributions of this paper can be summarized in the following points:
\begin{itemize}
\item The average number of packet transmissions required by a receiver to recover a generation has been expressed in closed form as a function of a preset deadline imposed on packet transmissions, without using Markov chain models as in \cite{Lucani2009_ITA, Lucani2009_Globecom} and \cite{Lucani2012}.
\item An upper bound on the average decoding delay at a receiver has been computed and shown to be tighter than the bounds presented in \cite{Eryilmaz2006} and \cite{Lucani2012}.
\item The delay analysis has covered both non-systematic and systematic random linear network coding for broadcast transmission, and has established that the systematic scheme incurs a shorter average decoding delay than non-systematic transmission when the generation size and the field size are small.
\item The proposed theoretical framework has been validated by a series of simulation results. The impact of the generation size, the field size and the deadline choice on the average decoding delay has also been explored.

\end{itemize}

The remainder of the paper has been structured as follows. Section~\ref{sec:system} describes the system model. Sections \ref{sec:NS-RLNC} and \ref{sec:S-RLNC} consider \mbox{non-systematic} and systematic network coding, respectively, and derive expressions for the average number of required coded packet transmissions and the average decoding delay. Section~\ref{sec:results} validates the proposed theoretical analysis and applies it to a practical setting that is based on the Long Term Evolution-Advanced (LTE-A) standard. Concluding remarks and future directions are summarized in Section~\ref{sec:conclusion}.


\section{System Model}
\label{sec:system}

We consider a system of one transmitter broadcasting coded packets to multiple receivers. For simplicity, a time step is set equal to the duration of a transmitted packet, implying that $K$ packets can be delivered to the receivers in $K$ time steps in perfect channel conditions. We define as $N\!=\!K+\Omega$ the predetermined number of coded packet transmissions per generation, where $\Omega$ denotes the permissible overhead, that is, the number of additional coded packet transmissions before the deadline expires. Once $K$ coded packets have been broadcast, each receiver is expected to recover the $K$ source packets in the subsequent $\Omega$ time steps.

A source packet $\mathbf{u}_i$, for $i=1,\dots,K$, is modeled as a sequence of $L$ symbols from a finite field of size $q$, that is, $\mathbf{u}_i\in\mathbb{F}^{L}_{q}$. At time step $j$, for $j=1,\dots,N$, the transmitter generates the coded packet  $\mathbf{x}_j\in\mathbb{F}^{L}_{q}$ as follows
\begin{equation} 
\mathbf{x}_j=\sum_{i=1}^{K}g_{i,j}\:\mathbf{u}_i\nonumber
\label{eq:single_step}
\end{equation}
and sends it over a broadcast channel characterized by packet erasure probability $\varepsilon$. The coefficients $g_{i,j}$ are selected uniformly at random from $\mathbb{F}_q$ \cite{Ho03}. If $n$ coded packets have been transmitted, the input to the broadcast channel can be written in matrix notation as 
\begin{equation} 
\left[\begin{array}{c} \mathbf{x}_{1}\\ \vdots\\ \mathbf{x}_{n} \end{array}\right]= \mathbf{G} 
\left[\begin{array}{c} \mathbf{u}_{1}\\ \vdots\\ \mathbf{u}_{K} \end{array}\right]
\label{eq:matrix_form} 
\end{equation}where the coefficients $g_{i,j}$ are the elements of $\mathbf{G}\in\mathbb{F}_{q}^{\,n\times K}$, which is referred to as the coding matrix. Let $\mathbf{y}_1,\ldots, \mathbf{y}_{m}$ denote the $m\leq n$ coded packets that a receiver successfully retrieved from the set of $n$ transmitted packets. Furthermore, let $\mathbf{M}$ represent the $m\times K$ decoding matrix constructed at the receiver from the $m$ rows of $\mathbf{G}$ that are associated with the retrieved packets. The relationship between $\mathbf{y}_1,\ldots, \mathbf{y}_{m}$ and the source packets $\mathbf{u}_1,\ldots, \mathbf{u}_{K}$ is 
\begin{equation} 
\left[\begin{array}{c} 
\mathbf{y}_{1}\\ \vdots\\ \mathbf{y}_{m} 
\end{array}\right]= \mathbf{M} 
\left[\begin{array}{c} \mathbf{u}_{1}\\ \vdots\\ \mathbf{u}_{K} 
\end{array}\right].\nonumber
\label{eq:after_detection}
\end{equation}The receiver can recover the $K$ source packets if and only if the rank of $\mathbf{M}$ is $K$, which implies that $\mathbf{M}$ contains a $K\times K$ invertible matrix.

\begin{figure}
    \centering
    	\includegraphics[width=0.65\columnwidth]{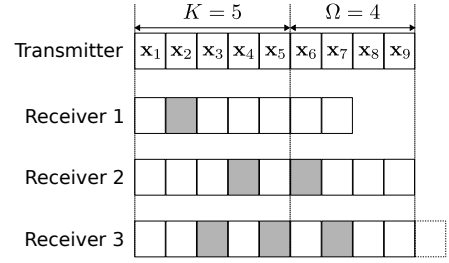}
	\vspace{1mm}
    \caption{A transmitter generates $N=9$ coded packets from a generation of $K=5$ source packets and broadcasts them to three receivers. The overhead is $\Omega=4$ packets. Grey slots depict erased packets at the respective receiver. Receivers 1 and 2 recover the generation after 7 and 9 packet transmissions, respectively. Receiver 3 fails to recover the generation by the set deadline.}
	\label{fig:system_model}
\end{figure}

The $n$ packet transmissions, which were sufficient for the recovery of the $K$ source packets of a particular generation by a receiver, can be expressed as $n=K+\omega$, where $\omega$ denotes the overhead for the generation under consideration. Fig.~\ref{fig:system_model} shows an example of three receivers, which attempt to recover a generation of $K=5$ source packets. Receiver~1 recovers the generation after $n=7$ packet transmissions and is not required to listen to the last two transmissions. Receiver~2 listens to all packet transmissions, i.e., $n=9$, and recovers the generation too. Receiver~3 needs one additional packet transmission after the deadline, which is shown as a dashed frame in Fig.~\ref{fig:system_model}, to reconstruct the $K=5$ source packets. Given that $N=9$ is a hard deadline, receiver 3 does not recover the generation and is in outage. Taking into account that $n$ and $\omega$ can be modeled as discrete random variables that represent packet transmissions but also elapsed time steps, the following definitions apply throughout this paper:

\begin{definition}
\label{def:outage}
The outage probability $P_\mathrm{out}$ is the probability that a receiver will not recover a generation by the set deadline of $N$ time steps, and is defined as \mbox{$P_\mathrm{out}=\mathrm{Pr}\left(n>N\right)$} or, equivalently, $P_\mathrm{out}=\mathrm{Pr}\left(\omega>\Omega\right)$.
\end{definition}

\begin{definition}
\label{def:avg_tx_packets}
The average number of packet transmissions $\bar{n}$ required by a receiver to recover a generation, provided that the deadline has not passed, is the expected value of $n$ for $n$ capped at $N$. It is defined as $\bar{n}=\mathrm{E}\left[n\right]$ for $K\leq n\leq N$, where $\mathrm{E}\left[\cdot\right]$ denotes the expectation operator. Similarly, the average overhead is defined as $\bar\omega=\mathrm{E}\left[\omega\right]$ for $0\leq \omega\leq \Omega$.
\end{definition}

\begin{definition}
\label{def:avg_delay}
The average decoding delay $\bar{d}$ at a receiver is the expected value of time steps (or, equivalently, the average number of packet transmissions) required by that receiver to recover a generation of $K$ source packets when no deadline is imposed. It is defined as $\bar{d}=\lim_{\Omega\rightarrow\infty}\bar{n}=K+\lim_{\Omega\rightarrow\infty}\bar{\omega}$.
\end{definition}

Based on these definitions, we understand that if the erasure probability $\varepsilon$ is small and the overhead $\Omega$ is sufficiently large, the outage probability $P_\mathrm{out}$ is close to zero, while $\bar{n}\approx\bar{d}$. As $\varepsilon$ increases, the value of $P_\mathrm{out}$ approaches~1, the value of $\bar{n}$ increases and eventually settles to $N$, and the value of $\bar{d}$ increases without bound. The system could reach a state where a receiver always listens to all of the $N$ coded packet transmissions ($\bar{n}=N$) but fails to recover the $K$ source packets of a generation with high probability. In that case, the deadline should be relaxed by increasing the value of $\Omega$ until $P_\mathrm{out}$ has been lowered to a desired value.

Two implementations of network coding are considered in the remainder of the paper, their outage behavior is studied, closed-form expressions for $\bar{n}$ are obtained, and tight bounds for $\bar{d}$ are derived.


\section{Non-systematic Network Coding}
\label{sec:NS-RLNC}

In conventional (non-systematic) random linear network coding, the $m \times K$ decoding matrix $\mathbf{M}$ will be a random matrix from $\mathbb{F}_q^{\,m\times K}$ and its rank will be $K$ with probability 
\begin{equation}
\begin{split}
\label{eq:full_rank}
P(m)&=\left\{
	\begin{array}{ll}
		\!\prod_{i=0}^{K-1}\left(1-q^{-m+i}\right),&\!\!\!\mbox{if }m \geq K\\[0.5em]
		\!0, &\!\!\!\mbox{if }m < K.
	\end{array}
\right.
\end{split}
\end{equation}
As mentioned in \cite{Trullols-Cruces2011}, $P(m)$ also represents the Cumulative Distribution Function (CDF) of the probability of receiving $K$ linearly independent coded packets, given the receipt of $m$ coded packets.

To compute the probability that a receiver will recover the $K$ source packets of a generation in $N=K+\Omega$ or fewer time steps, the probability that a generation will be recovered in \textit{exactly} $n=K+\omega$ steps needs to be obtained first. Let $s^{n}_\rho$ denote the following statement:
\begin{equation}
s^{n}_\rho: \textrm{The decoding matrix}\;\mathbf{M}\;\textrm{has rank}\;\rho\;\textrm{at time step}\;n,\nonumber
\end{equation}
the desired Probability Mass Function (PMF) of $\omega$, denoted by $f(\omega)$, can be expressed as the product of two terms:
\begin{equation}
f(\omega) =\Pr\bigl(s^{K+\omega}_{K}\;|\;s^{K+\omega-1}_{K-1}\bigr)\,\Pr\bigl(s^{K+\omega-1}_{K-1}\bigr).
\label{eq:PMF_product}
\end{equation}The first term of the product in \eqref{eq:PMF_product} considers the case when the receiver has already collected $K-1$ linearly independent coded packets in $K+\omega-1$ time steps. It corresponds to the probability that the coded packet transmitted at time step $K+\omega$ has been successfully delivered to the receiver with probability $(1-\varepsilon)$ and is the $K$-th required linearly independent coded packet with probability $p_K$. We can thus write
\begin{equation}
\Pr\bigl(s^{K+\omega}_{K}\;|\;s^{K+\omega-1}_{K-1}\bigr)=(1-\varepsilon)\,p_K.
\label{eq:first_product_term}
\end{equation}
The second term of the product in \eqref{eq:PMF_product} represents the probability that $K-1$ linearly independent coded packets have been recovered in the first $K+\omega-1$ time steps. If we denote by $P_r(m)$ the probability that a random matrix over $\mathbb{F}_q^{m\times K}$ has rank $r$, where $0\leq r\leq\min(m,K)$, the second term in \eqref{eq:PMF_product} assumes the following form
\begin{equation}
\displaystyle\Pr\bigl(s^{K+\omega-1}_{K-1}\bigr)\!=\!\!\!\!\sum_{m=K\negsp-\negsp1}^{\tau_\omega}\!\!\!{\tau_\omega \choose m}\varepsilon^{\tau_\omega-m}(1-\varepsilon)^{m}\negsp\negsp P_{K\negsp-\negsp1}(m)
\label{eq:second_product_term}
\end{equation}
where we set $\tau_\omega=K+\omega-1$ for compactness of notation. A recursive relationship that links $p_K$ in \eqref{eq:first_product_term} and $P_{K\negsp-\negsp1}(m)$ in \eqref{eq:second_product_term} with the well-defined probability $P(m)$ in \eqref{eq:full_rank} is
\begin{equation}
P(m\negsp+\negsp1)=P_{K\negsp-\negsp1}(m)\,p_K+P(m).
\label{eq:recursive_full_rank}
\end{equation}
In other words, a full-rank $(m+1)\times K$ random matrix can be obtained from an $m\times K$ random matrix, if the $m\times K$ matrix has rank $K\negsp-\negsp1$ and a linearly independent row is appended to it or the $m\times K$ matrix has already rank $K$. Note that $P(K)=P_{K\negsp-\negsp1}(K\negsp-\negsp1)\,p_K$ for $m=K-1$. Substituting \eqref{eq:first_product_term} and \eqref{eq:second_product_term} into \eqref{eq:PMF_product}, and using \eqref{eq:recursive_full_rank}, gives 
\begin{equation}
\begin{split}
f(\omega)&={\tau_\omega \choose K\,\shortminus\,1}\,\varepsilon^{\omega}\,(1\,\shortminus\,\varepsilon)^{K}\,P(K)\,+\\
&+\!\!\sum_{m=K}^{\tau_\omega}\!\!{\tau_\omega \choose m}\varepsilon^{\tau_\omega-m\,}\negsp\negsp(1\,\shortminus\,\varepsilon)^{m+1\,}\!\bigl[P(m\!+\!1)\,\shortminus\, P(m)\bigr]\negsp.
\end{split}
\label{eq:PMF_final}
\end{equation}
The CDF of $\omega$, denoted by $F(\omega)$, describes the probability that the considered receiver will recover the $K$ source packets of a generation in \textit{up to} $n=K+\omega$ time steps and is equal to $F(\omega)=\sum_{i=0}^{\omega}f(i)$.

\begin{remark}
Assume that $K-1$ coded packets have been collected by the receiver and a $(K-1)\times K$ decoding matrix $\mathbf{M}$ has been constructed. Each time $\mathbf{M}$ is augmented by a row due to the successful delivery of an additional coded packet, the receiver computes its rank. If the rank of $\mathbf{M}$ is $K$ when its dimensions are $m\times K$, we can conclude that $m-K+1$ rank checks have been carried out. Regular rank checks ensure that the $K$ source packets will be recovered as soon as $K$ linearly independent coded packets are received but do not affect the probability of the $m\times K$ decoding matrix $\mathbf{M}$ having rank $K$. Based on this observation, the CDF of $\omega$ should be given by
\begin{equation}
F(\omega)\!=\!\!\sum_{m=K}^{K+\omega}\!\!{K+\omega\, \choose m}\varepsilon^{K+\omega-m}(1-\varepsilon)^{m}P(m)
\label{eq:CDF_d}
\end{equation}
and the PMF of $\omega$ could be obtained from $F(\omega)$ as follows:
\begin{equation}
f(\omega)=F(\omega)-F(\omega\negsp\negsp-\negsp\negsp1).
\label{eq:PMF_d_equiv}
\end{equation}
Expression \eqref{eq:PMF_d_equiv}, which has been invoked in the literature, e.g.,\cite{Khan2015}, is indeed equivalent to \eqref{eq:PMF_final} and Appendix~\ref{apdx:equiv_PMF} describes the steps for obtaining \eqref{eq:PMF_d_equiv} from \eqref{eq:PMF_final}. The probability analysis that follows rely on both \eqref{eq:PMF_final} and \eqref{eq:CDF_d}.
\end{remark}

Given that $n\negsp=\negsp K\negsp+\negsp \omega$, the PMF and CDF of the overhead $\omega$ also describe the probability distribution of the required packet transmissions $n$, if $K\negsp+\negsp \omega$ is replaced by $n$ in \eqref{eq:PMF_final} and \eqref{eq:CDF_d}. The probability that a receiver will fail to recover the $K$ source packets in $N=K+\Omega$ time steps represents the outage probability and is given by $P_\mathrm{out}=1-F(\Omega)$ or equivalently
\begin{equation}
P_\mathrm{out}=1-\sum_{m=K}^{N}\!\!{N \choose m}\varepsilon^{N-m}(1-\varepsilon)^{m}P(m).
\label{eq:NC_dec_delay}
\end{equation}
Calculation of the average number of packet transmissions $\bar{n}$ and derivation of bounds for the average decoding delay $\bar{d}$ are more involved and are described in detail in the following proposition and corollary.

\begin{proposition}
\label{pro:avg_buf_delay}
If a transmitter broadcasts $N=K+\Omega$ coded packets over a channel subjected to packet erasures with probability $\varepsilon$, the average number of coded packet transmissions that a receiver is required to listen to, so that it stands the best chance of recovering the $K$ source packets, is
\begin{equation}
\bar{n} = N - \sum\limits_{\nu=0}^{\Omega-1}\sum_{m=K}^{K+\nu}\!\!{K+\nu\, \choose m}\varepsilon^{K+\nu-m}(1-\varepsilon)^{m}P(m)
\label{eq:NC_dec_delay}
\end{equation}
where $\Omega$ represents the permissible overhead.
\end{proposition}
\begin{IEEEproof}
The average overhead $\bar{\omega}$ signifies the average number of coded packet transmissions that a receiver is required to listen to, in addition to the $K$ original coded packet transmissions. A receiver will be interested in the coded packet transmitted at time step $K+1$ if $K$ linearly coded packets have not been recovered in the previous $K$ time steps with probability $1-F(0)$. In general, the considered receiver will be interested in the coded packet transmitted at time step $K+1+\nu$ if it has failed to collect $K$ linearly independent coded packets in the previous $K+\nu$ time steps with probability $1-F(\nu)$, for $0\leq \nu\leq \Omega-1$. Adding up the probabilities that weight each coded packet transmission for $\nu=0,\ldots,\Omega-1$ gives the average overhead, that is
\begin{equation}
\bar{\omega} = \Omega - \sum\limits_{\nu=0}^{\Omega-1}F(\nu).
\label{eq:delay_NC_fundamental}
\end{equation}
The average number of coded packet transmissions $\bar{n}$ can be obtained from \eqref{eq:delay_NC_fundamental} if $K$ is added to the right-hand side of \eqref{eq:delay_NC_fundamental} and $F(\nu)$ is expanded using \eqref{eq:CDF_d}.
\end{IEEEproof}

\begin{figure}[t]
	\centering
	\includegraphics[width=1\columnwidth]{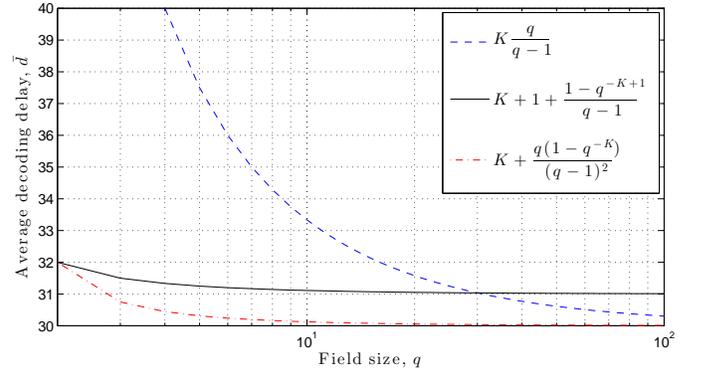}
	\caption{Comparison of the upper bounds \eqref{eq:tight_bound}, \eqref{eq:bound_1} and \eqref{eq:bound_2} for $K=30$.}
	\label{fig:compare_bounds}
\end{figure}

\begin{corollary}
\label{pro:avg_buf_delay_unconstrained}
Consider a transmitter that employs random linear network coding over $\mathbb{F}_q$ on $K$ source packets and generates coded packets. If a potentially infinite number of coded packets can be transmitted over a broadcast channel to multiple receivers, the average decoding delay $\bar{d}$ incurred by each receiver can be bounded as follows:
\begin{equation}
\frac{K}{1-\varepsilon}\leq\bar{d}<\frac{1}{1-\varepsilon}\left[K+\frac{q\left(1-q^{-K}\right)}{(q-1)^2}\right]
\label{eq:NS_decodingdelay_bounds}
\end{equation}
where $\varepsilon$ is the packet erasure probability of the broadcast channel.
\end{corollary}
\begin{IEEEproof}
The lower bound can be easily understood if we take into account that, out of the $n$ transmitted coded packets, $(1-\varepsilon)n$ coded packets will be delivered on average to each receiver. A receiver has a chance of recovering the generation of $K$ source packets only if $(1-\varepsilon)n$ is at least equal to $K$. Derivation of the upper bound is nontrivial and a detailed proof is provided in Appendix \ref{apdx:proof_upper_bound}.
\end{IEEEproof}

\vspace{1mm}

\begin{remark}
The minimum of two upper bounds on $\bar{d}$ has been proposed in \cite[Lemma 2]{Lucani2012}. In particular, the authors of \cite{Lucani2012} proved that
\begin{equation}
\bar{d}<\frac{1}{1-\varepsilon}\min\negsp\negsp\left\{K\frac{q}{q-1},\, K+1+\frac{1-q^{-K+1}}{q-1}\right\}.
\label{eq:Lucani_upperbound}
\end{equation}
Given that $K\geq2$ and $q\geq2$ in random linear network coding, we can show that the upper bound in \eqref{eq:NS_decodingdelay_bounds} is tighter than the two bounds in \eqref{eq:Lucani_upperbound}. Factor $1/(1-\varepsilon)$ appears in all bounds and can be omitted from this comparison. The upper bound in \eqref{eq:NS_decodingdelay_bounds} can thus assume the form
\begin{equation}
K + \frac{q\left(1-q^{-K}\right)}{(q-1)^2} = K + \left(\frac{1}{q-1}\right)\!\negsp\left(\frac{q}{q-1}\right)\!\negsp\left(1-q^{-K}\right)
\label{eq:tight_bound}
\end{equation}
whilst the first and second bounds in \eqref{eq:Lucani_upperbound} can be rewritten as:
\begin{align}
\label{eq:bound_1}
K\frac{q}{q-1}&=K+\left(\frac{1}{q-1}\right)\!K\\
\label{eq:bound_2}
K+1+\frac{1\negsp-\negsp q^{-K+1}}{q\negsp-\negsp1}&=K + \left(\frac{q}{q-1}\right)\!\negsp\left(1-q^{-K}\right)
\end{align}
respectively. The product terms on the right-hand side of \eqref{eq:tight_bound}, \eqref{eq:bound_1} and \eqref{eq:bound_2} are all positive numbers. We also know that $q/(q-1)\negsp\leq\negsp K$ and \mbox{$0.75\negsp\leq\negsp(1- q^{-K})\negsp<\negsp 1$}, therefore \eqref{eq:tight_bound} is a lower bound of \eqref{eq:bound_1}. On the other hand, \mbox{$1/(q-1)\leq 1$} hence \eqref{eq:tight_bound} bounds \eqref{eq:bound_2} from below too. A graphical demonstration of the tightness of \eqref{eq:tight_bound} is shown in Fig.~\ref{fig:compare_bounds}. This concludes the proof that \eqref{eq:NS_decodingdelay_bounds} is a tighter bound than \eqref{eq:Lucani_upperbound}.
\end{remark}


\section{Systematic Network Coding}
\label{sec:S-RLNC}

In systematic network coding, the transmitter broadcasts $N$ packets; the first $K$ packets are identical to the $K$ source packets of a generation and are referred to as \textit{systematic} packets, while the remaining \mbox{$\Omega=N-K$} packets are generated as in the non-systematic case. At time step $n^{\prime}$, the transmitter has broadcast $n^{\prime}=K+\omega^{\prime}$ packets, where the overhead consists of $\omega^\prime$ coded packets. The $n^\prime \times K$ coding matrix $\mathbf{G}$ has the following structure
\setlength{\arraycolsep}{0.5em}
\begin{equation}
\mathbf{G} =
\left[
\begin{array}{c}
 \mathbf{I}\\
 \mathbf{C}
\end{array}
\right]
\label{eq:sys_G}
\end{equation}
where $\mathbf{I}$ is the $K \times K$ identity matrix and $\mathbf{C}$ is a random matrix from $\mathbb{F}_q^{\omega^\prime \times K}$. If a receiver successfully collects $m$ packets, of which $h$ are systematic and the remaining $m-h$ are coded, the  $m \times K$ decoding matrix $\mathbf{M}$ will be constructed. The probability that $\mathbf{M}$ has rank $K$ for $m\geq K$ is given by
\begin{equation}
P^{\prime}(m,\negsp\omega^{\prime})\!=\!\frac{{\displaystyle\sum_{h = h_{\min}}^{K}}\!\!\!\displaystyle\binom{K}{h}\!\binom{\omega^{\prime}}{m\negsp-\negsp h}\!\displaystyle{\prod_{i=0}^{K-h-1}}\!\!\!\!\left(1-q^{-m+h+i}\right)}{\displaystyle\binom{K+\omega^{\prime}}{m}}
\label{eq:sys_full_rank}
\end{equation}
where $h_{\min}=\max{(0,\,m\negsp-\negsp\omega^{\prime})}$ \cite{Shrader09}, \cite[Lemma 1]{Jones15}. For \mbox{$K\negsp-\negsp h\negsp-\negsp 1\negsp<\negsp 0$}, the product in the right-hand side of \eqref{eq:sys_full_rank} becomes an empty product and is equal to 1.

The probability of $\mathbf{M}$ having rank $K$ is higher for systematic network coding than for non-systematic network coding\footnote{This is a property of random linear network coding and cannot be generalized to channel coding at the physical layer.} when arithmetic operations are over $\mathbb{F}_2$ \cite[Proposition 2]{Jones15}, that is, $P^{\prime}(m,\omega^{\prime})>P(m)$ for $q=2$. This proposition is generalized for any valid value of $q$ in the following lemma.
\begin{lemma}
\label{pro:S_better_than_NS}
Consider the case where systematic network coding is used and only $\omega^\prime$ of the $K+\omega^\prime$ transmitted packets are coded, and the case where non-systematic network coding is employed and all of the $K+\omega^\prime$ transmitted packets are coded. In both cases, the number of source packets is $K$ and the number of packets delivered to a receiver is $m$. If $P^\prime(m,\omega^\prime)$ and $P(m)$ are the probabilities that the decoding matrix in each case has full rank, then
\begin{equation}
P^{\prime}(m,\negsp \omega^{\prime})>P(m)
\label{eq:SvsNS_inequality}
\end{equation}
for $K\leq m \leq K\negsp+\negsp \omega^\prime$ and $q\geq 2$.
\end{lemma}
\begin{IEEEproof}
Using \eqref{eq:full_rank}, we first look at the ratio:
\begin{equation}
\begin{split}
\frac{P(m)}{\displaystyle\prod_{i=0}^{K-h-1}\!\!\!\!\left(1-q^{-m+h+i}\right)} = &\prod_{i = 0}^{K-h-1}\!\!\frac{\left(1-q^{-m+i}\right)}{\left(1-q^{-m+h+i}\right)}\,\times\\
&\times\prod_{i = K-h}^{K-1}\!\!\left(1-q^{-m+i}\right).
\end{split}
\label{eq:ns_ratio}
\end{equation}Both products of terms in the right-hand side of \eqref{eq:ns_ratio} generate values that are smaller than~$1$, therefore
\begin{equation}
\displaystyle\prod_{i=0}^{K-h-1}\!\!\!\!\left(1-q^{-m+h+i}\right) > P(m)
\label{eq:ns_inequality}
\end{equation}
for $m\geq K$ and $0<h\leq K$. Based on the Chu-Vandermonde identity~\cite[p. 41]{Koepf98} and the Vandermonde's convolution~\cite[p. 29]{Roman84}, the following relation holds
\begin{equation}
\sum_{h = h_{\min}}^{K}\!\!\binom{K}{h} \binom{\omega^\prime}{m-h} = \binom{K+\omega^\prime\,}{m}
\label{eq.vander}
\end{equation}
where $h_{\min}=\max{(0,\,m\negsp-\negsp\omega^{\prime})}$. Using both \eqref{eq:ns_inequality} and \eqref{eq.vander}, the series below can be bounded as follows:
\setlength{\arraycolsep}{0.0em}
\begin{eqnarray}
&\displaystyle\sum_{h = h_{\min}}^{K}\!\!\!\displaystyle\binom{K}{h}\!\binom{\omega^{\prime}}{m\negsp-\negsp h}\!&\displaystyle{\prod_{i=0}^{K-h-1}}\!\!\!\!\left(1-q^{-m+h+i}\right)>\nonumber\\
&&>\displaystyle P(m)\sum_{h = h_{\min}}^{K}\!\!\!\binom{K}{h} \binom{\omega^\prime}{m-h}\nonumber\\
&&=\displaystyle P(m)\,\binom{K+\omega^\prime\,}{m}.
\label{eq:lemma_working}
\end{eqnarray}
\setlength{\arraycolsep}{5pt}
Combining \eqref{eq:lemma_working} with \eqref{eq:sys_full_rank} yields the desired \eqref{eq:SvsNS_inequality}.
\end{IEEEproof}

\vspace{1mm}

The number of packet transmissions $n^\prime$ and the overhead $\omega^\prime$ can be modeled as discrete random variables. Following the same line of thought and reasoning as in Section \ref{sec:NS-RLNC}, the PMF and CDF of $\omega^\prime$ assume the form
\begin{equation}
\begin{split}
&f^{\prime}(\omega^{\prime})={\tau_{\omega^{\prime}} \choose K-1}\,\varepsilon^{\omega^{\prime}}\,(1-\varepsilon)^{K}\,P^{\prime}(K,\omega^{\prime})\,+\\
&\!+\!\sum_{m=K}^{\tau_{\omega^{\prime}}}\!\!{\tau_{\omega^{\prime}} \choose m}\varepsilon^{\tau_{\omega^{\prime}}-m\,}\negsp\negsp(1-\varepsilon)^{m+1\,}\!\bigl[P^{\prime}\negsp(m\!+\!1,\omega^{\prime})-P^{\prime}\negsp(m,\omega^{\prime})\bigr]\nonumber
\end{split}
\label{eq:sys_PMF_final}
\end{equation}
and
\begin{equation}
F^{\prime}(\omega^{\prime})=\sum_{m=K}^{K+\omega^{\prime}}\!\!{K+\omega^{\prime}\, \choose m}\varepsilon^{K+\omega^{\prime}-m}(1-\varepsilon)^{m}P^{\prime}(m,\omega^{\prime})\nonumber
\label{eq:sys_CDF_d}
\end{equation}
respectively, where $\tau_{\omega^\prime}=K+\omega^{\prime}-1$. Similarly to non-systematic network coding, the outage probability of systematic network coding can be computed using $P^{\prime}_\mathrm{out}=1-F^{\prime}(\Omega)$, while the average number of required packet transmissions $\bar{n}^{\prime}$ and the average decoding delay $\bar{d}^{\prime}$ are obtained in the following proposition and corollaries.
\begin{proposition}
\label{pro:sys_avg_buf_delay}
For each generation, a transmitter sends $K$ source packets followed by $\Omega$ coded packets over a channel subjected to packet erasures with probability $\varepsilon$. The average number of packet transmissions that a receiver is required to listen to, so that it stands the best chance of recovering the $K$ source packets of a generation, is
\begin{equation}
\bar{n}^{\prime}\negsp\!=\! N\negsp\negsp -\negsp\negsp \sum\limits_{\nu=0}^{\Omega-1}\negsp\sum_{m=K}^{K+\nu}\!\!\negsp{K\negsp\negsp+\negsp\negsp\nu \choose m}\varepsilon^{K+\nu-m}(1\negsp-\negsp\varepsilon)^{m}P^{\prime}\negsp(m,\nu).
\label{eq:SNC_dec_delay}
\end{equation}
\end{proposition}
\begin{IEEEproof}
The same steps as in Proposition \ref{pro:avg_buf_delay} can be followed to first obtain $\bar{\omega}^{\prime}$ as a function of $F^{\prime}(\nu)$, for $0\leq\nu\leq\Omega-1$, and then use $\bar{n}^{\prime}=K+\bar{\omega}^{\prime}$ to derive \eqref{eq:SNC_dec_delay}.
\end{IEEEproof}

\begin{corollary}
In a network-coded broadcast system, which uses a given field size $q$, generation size $K$ and deadline values $N$ and $\Omega$, a receiver is required to listen to fewer packet transmissions, on average, when systematic network coding is employed as opposed to non-systematic network coding.
\label{pro:SvsNS_avg_delay}
\end{corollary}
\begin{IEEEproof}
According to Lemma \ref{pro:S_better_than_NS}, $P^{\prime}(m,\omega^{\prime})>P(m)$ holds. Therefore, $F^{\prime}(\omega^{\prime})>F(\omega)$ for $\omega^{\prime}=\omega$ also holds, which leads to the conclusion that $\bar{n}^{\prime}<\bar{n}.$
\end{IEEEproof}

\begin{figure}[t]
	\centering
	\includegraphics[width=1\columnwidth]{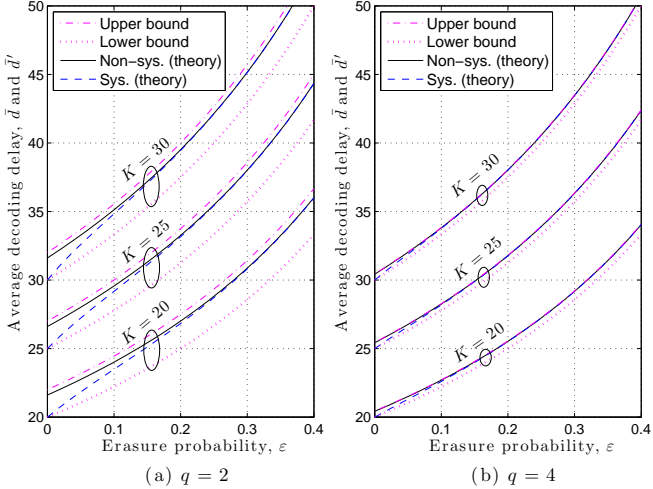}
	\caption{The upper and lower bounds, presented in \eqref{eq:NS_decodingdelay_bounds}, are compared with the theoretical average decoding delay of non-systematic (non-sys.) and systematic (sys.) network coding. Finite fields of size (a) $q\!=\!2$ and (b) $q\!=\!4$ are considered. Results for generation sizes $K\!\in\!\{20,25,30\}$ are displayed.}
	\label{fig:cor1_and_cor3}
\end{figure}

\begin{corollary}
\label{pro:SvsNS_avg_delay_infty}
If systematic network coding is used to transmit a potentially infinite number of packets over the broadcast erasure channel, the average decoding delay $\bar{d}^{\prime}$ is confined within the same bounds as in the non-systematic case. 
\end{corollary}
\begin{IEEEproof}
The proof follows from Corollary \ref{pro:SvsNS_avg_delay}. Given that $\bar{n}^{\prime}<\bar{n}$ for any value of $\Omega$, the bounds used in \eqref{eq:NS_decodingdelay_bounds} can also be used to bound $\bar{d}^{\prime}=\lim_{\Omega\rightarrow\infty}\bar{n}^{\prime}$. The tightness of the bounds is demonstrated in Fig.~\ref{fig:cor1_and_cor3}.
\end{IEEEproof}

\vspace{1mm}

\begin{remark}
For large values of $q$, the first $K$ received packets will be linearly independent with high probability for both non-systematic and systematic random linear network coding. Indeed, if $q\!\rightarrow\!\infty$, we obtain $P(m)\!=\!P^{\prime}(m,\omega^\prime)\!=\!1$ for $m\!\geq\!K$, hence, the average number of packet transmissions for both implementations of network coding will be
\begin{equation}
\bar{n} = \bar{n}^{\prime} = N - \sum\limits_{\nu=0}^{\Omega-1}\sum_{m=K}^{K+\nu}\!\!{K+\nu\, \choose m}\varepsilon^{K+\nu-m}(1-\varepsilon)^{m}
\label{eq:NC_dec_delay}
\end{equation}
while the average decoding delay will be $\bar{d}=\bar{d}^{\prime}=K/(1-\varepsilon)$. In practice, as shown in Fig.~\ref{fig:cor1_and_cor3} and discussed in Section~\ref{sec:results}, non-systematic and systematic random linear network coding have similar delay performances for $q\geq 4$.
\end{remark}

\begin{figure}[t]
	\centering
	\includegraphics[width=1\columnwidth]{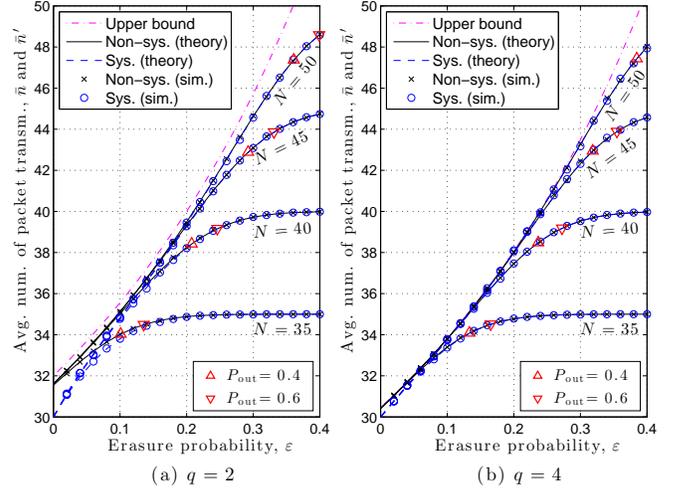}
	\caption{Comparison of results obtained from theoretical expressions (theory) and through simulations (sim.) for non-systematic (non-sys.) and systematic (sys.) network coding. The generation size is set to $K=30$, while finite fields of size (a) $q=2$ and (b) $q=4$ are considered.}
	\label{fig:validate_theory}
\end{figure}


\section{Numerical Results}
\label{sec:results}

In Sections \ref{sec:NS-RLNC} and \ref{sec:S-RLNC}, exact expressions for the average number of required packet transmissions and bounds for the average decoding delay at a receiver were obtained for non-systematic and systematic random linear network coding. This section validates the  proposed theoretical framework, reflects on the delay performance of the two implementations of network coding for broadcast transmission and extends the framework to a particular LTE-A example case.

\subsection{Assessment of the Proposed Framework}
\label{sec:val_theory}

A comparison between theoretical values and simulation results is presented in Fig.~\ref{fig:validate_theory}. Specifically, the average number of required packet transmissions was computed through Monte Carlo simulations for generations of $K=30$ source packets and different values of total packet transmissions $N$ per generation, packet erasure probability $\varepsilon$ and field size $q$. We observe that simulation results coincide with theoretical evaluations based on \eqref{eq:NC_dec_delay} for non-systematic network coding, and \eqref{eq:SNC_dec_delay} for systematic network coding. The upper bound has been obtained using \eqref{eq:NS_decodingdelay_bounds} for both schemes. We notice that the average number of required packet transmissions closely follows the upper bound on the decoding delay for as long as the erasure probability $\varepsilon$ is low enough or the total number of transmitted packets $N$ is sufficiently large for a receiver to collect $K$ linearly independent coded packets. Otherwise, the average number of required packet transmissions deviates from the upper bound and converges to the total number of packet transmissions. When $\bar{n}$ (or $\bar{n}^{\prime}$) is equal to $N$, a receiver is required to listen to all packet transmissions but the outage probability is markedly high, as can be inferred from Fig.~\ref{fig:validate_theory}.

\begin{figure}[t]
	\centering
	\includegraphics[width=1\columnwidth]{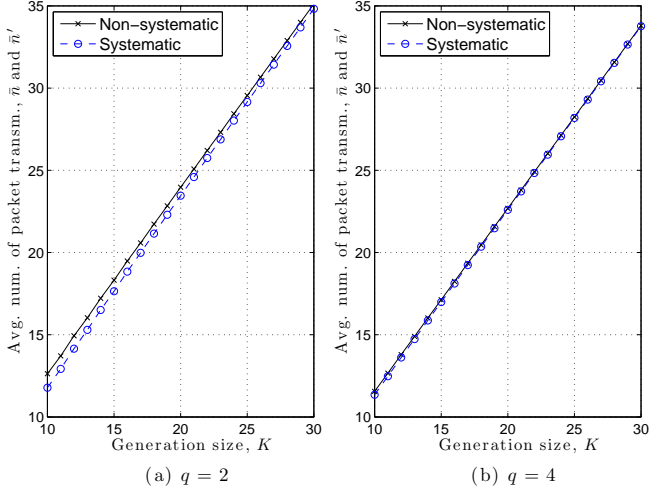}
	\caption{Average number of required packet transmissions for non-systematic and systematic network coding, when (a) $q=2$ and (b) $q=4$. The generation size $K$ takes values in $\{10,\ldots,30\}$, the total number of transmitted packets is $N=\lfloor 1.5K \rfloor$ and the erasure probability is $\varepsilon=0.1$.}
	\label{fig:nonsys_vs_sys}
\end{figure}

Having confirmed the validity and accuracy of the derived expressions, we shall now study them in more detail to gain insight into the interplay between system parameters. Delay performance curves for non-systematic and systematic network coding, obtained from \eqref{eq:NC_dec_delay} and \eqref{eq:SNC_dec_delay} respectively, are compared in Fig.~\ref{fig:nonsys_vs_sys}. The generation size $K$ ranges from $10$ to $30$ source packets. The total number of transmitted packets has been set to $N=\lfloor 1.5K \rfloor$, where $\lfloor\cdot\rfloor$ denotes the integer part of a number. The reduced decoding complexity of systematic network coding, which has been reported in \cite{Lucani2012}, is complemented by a decreased number of required packet transmissions or, equivalently, a smaller delay in decoding a generation when the chosen field size is $q=2$, as shown in Fig.~\ref{fig:nonsys_vs_sys}a. As the generation size increases, the delay performance of systematic network coding becomes comparable to that of non-systematic network coding. Fig.~\ref{fig:nonsys_vs_sys}b shows that, when the finite field consists of four (or more) elements, the delay profiles of the two schemes are nearly identical.

\begin{figure}[t]
	\centering
	\includegraphics[width=1\columnwidth]{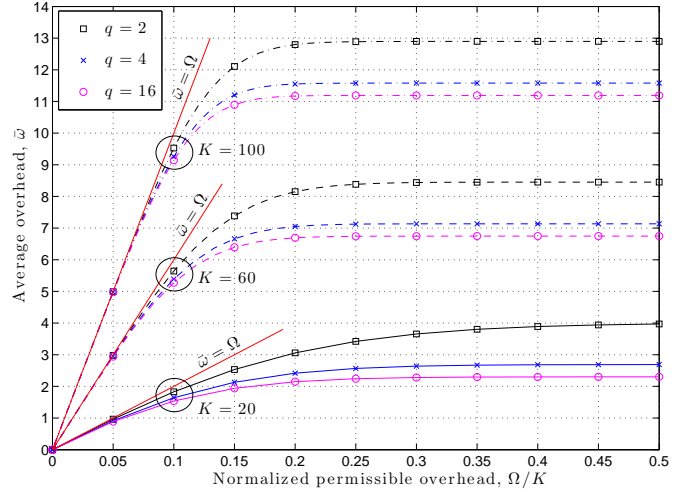}
	\caption{Average overhead $\bar{\omega}$ as a function of the permissible overhead $\Omega$, which has been expressed as a fraction of the generation size $K$. The erasure probability is $\varepsilon=0.1$, while the generation size and the field size take values in $K\in\{20,60,100\}$ and $q\in\{2,4,16\}$, respectively.}
	\label{fig:avgVSmax_overheads}	
\end{figure}

For non-systematic network coding, the average overhead $\bar{\omega}$, which is given in \eqref{eq:delay_NC_fundamental}, for an increasing ratio $\Omega/K$, erasure probability $\varepsilon=0.1$ and various values of $K$ and $q$, is plotted in Fig.~\ref{fig:avgVSmax_overheads}. The ratio $\Omega/K$ represents the number of additional coded packet transmissions expressed as a fraction of the considered generation size. Let us focus on the case of a transmitter that encodes generations of \mbox{$K=60$} source packets and broadcasts $60$ coded packets followed by $\Omega$ additional coded packets for each generation, where $\Omega\in\{0,\dots,30\}$. The line described by equation $\bar{\omega}=\Omega$ in Fig.~\ref{fig:avgVSmax_overheads} for $K=60$ depicts the worst-case average overhead, which occurs when the outage probability is high and a receiver is driven to listen to all of the packet transmissions until the deadline expires. Regardless of the size $q$ of the finite field, if \mbox{$\Omega<3$}, or equivalently \mbox{$\Omega/K<0.05$}, packet erasures prevent a receiver from collecting a sufficient number of coded packets and retrieving the source packets. On the other hand, if \mbox{$D\geq18$} or \mbox{$D/K\geq0.3$}, the average overhead $\bar{\omega}$ stabilizes at $8.45$, $7.13$ and $6.74$ time steps for $q$ equal to $2$, $4$ and $16$, respectively. Based on \eqref{eq:upper_bound}, the corresponding upper bounds (not shown in Fig.~\ref{fig:avgVSmax_overheads}) are $8.88$, $7.16$ and $6.74$. We conclude that if the preassigned value of $\Omega$ is increased beyond $18$ coded packets when $K=60$ and $\varepsilon=0.1$, the extra coded packet transmissions will be wasteful because they will have no impact on the average decoding delay at a receiver. Furthermore, we observe that the cost of adopting field $\mathbb{F}_2$ over $\mathbb{F}_4$ and $\mathbb{F}_{16}$ is the transmission of $1.32$ and $1.71$ extra coded packets, on average, respectively. However, $\mathbb{F}_2$ has the advantage of the least computationally expensive encoding and decoding processes. This observation is reinforced if we consider the range \mbox{$3\leq \Omega<18$} or \mbox{$0.05\leq \Omega/K<0.3$} in Fig.~\ref{fig:avgVSmax_overheads} and notice that the differences in performance are markedly smaller than those for higher values of $\Omega$. Similar trends can also be noted for $K=20$ and $K=100$ in Fig.~\ref{fig:avgVSmax_overheads}.

In summary, Fig.~\ref{fig:validate_theory} demonstrated that, for a given total number of coded packet transmissions $N$, the exact expressions for the average number of required packet transmissions $\bar{n}$ can be used to identify the erasure probability for which $\bar{n}$ deviates from the upper bound on the decoding delay $\bar{d}$ and saturates. Fig.~\ref{fig:nonsys_vs_sys} established that the decoding delay can be reduced for small generation sizes, if systematic network coding is selected over its non-systematic counterpart when arithmetic operations are over $\mathbb{F}_2$. Finally, Fig.~\ref{fig:avgVSmax_overheads} showed that, for a given erasure probability, the theoretical framework can determine the range from which values for the permissible overhead $\Omega$ can be drawn; outside this range, the amount of transmitted coded packets is either insufficient or excessive. 


\subsection{Performance Evaluation in an LTE-A System}
\label{sec:LTE}

\begin{figure}[t]
	\centering
	\includegraphics[width=1\columnwidth]{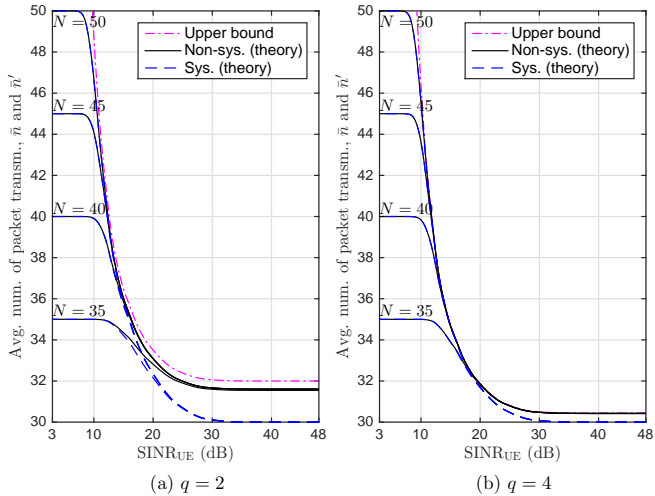}
	\caption{Average number of TB transmissions required as a function of the SINR at the User Equipment (UE). The generation size is $K=30$, while (a) $q=2$ and (b) $q=4$. The MCS with index $13$ has been used (16-QAM).}
	\label{fig:nnp}
\end{figure}

In order to investigate the delay performance of non-systematic and systematic random linear network coding for broadcast services in a practical setting, we consider an LTE-A system that consists of $19$ base stations equipped with three-sector antennas; $18$ of the base stations are evenly distributed on the circumference of three nested rings centered at the reference base station, which broadcasts a network-coded stream of packets. The inter-site distance of the base stations is $500$~m. Our simulations are based on the physical layer parameters prescribed in the 3GPP's benchmark Case~$1$~\cite{TR_36_814}.

The downlink phase of LTE-A relies on Orthogonal Frequency Division Multiple Access (OFDMA). As such, radio resources can be modeled as a time-frequency grid. In order to meet the physical layer constraints of LTE-A, each Transport Block (TB) carries one packet, either systematic or coded. Each TB spans a variable bandwidth and lasts for $10$~ms, i.e., it occupies $12$ OFDM symbols. In our simulations, we set the TB bandwidth equal to 1.62~MHz, i.e., we refer to TBs with a bandwidth equal to three \mbox{LTE-A} resource blocks~\cite{TR_36_814}. For simplicity, the radio resource mapping imposes that only one TB can be transmitted every 10~ms. Hence, a direct relationship between the number of packet transmissions $n$ or $n^\prime$ and the transmission time  has been established. We employed a fixed Modulation and Coding Scheme (MCS) for the transmission of each TB. The adopted MCS determines the TB error probability, which coincides with the erasure probability $\varepsilon$ in the considered setup. Further details on the LTE-A simulator can be found in~\cite[Section V]{Tassi2015}.

\begin{figure}[t]
	\centering
	\includegraphics[width=1\columnwidth]{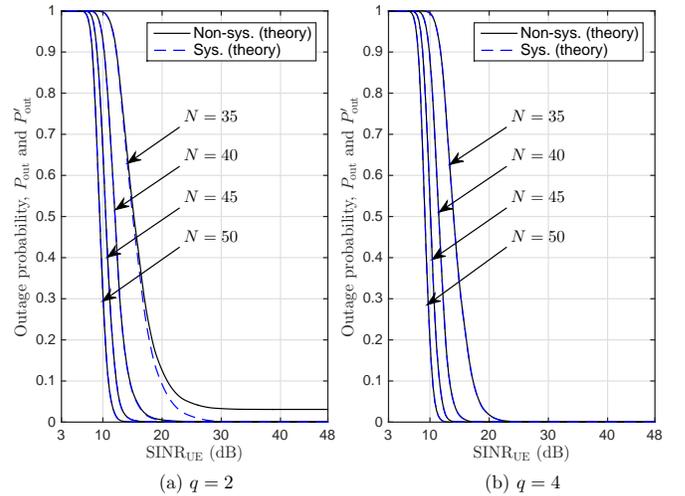} 
	\caption{Outage probability as a function of the SINR at the User Equipment (UE). The generation size is set to $K=30$, while (a) $q=2$ and (b) $q=4$. The MCS with index $13$ has been used (16-QAM).}
	\label{fig:opp}
\end{figure}

A user that receives the transmitted packet stream was considered. Fig.~\ref{fig:nnp} shows the values of $n$ and $n^\prime$ as a function of the user Signal-to-Interference-plus-Noise Ratio (SINR), while Fig.~\ref{fig:opp} depicts the outage probabilities $P_\mathrm{out}$ and $P^{\prime}_\mathrm{out}$ in terms of the user SINR. In both cases, the field size is $q\in\{2,4\}$ and the index of the adopted MCS is $13$~\cite{sesia2011lte}, i.e., $16$-QAM having a spectral efficiency of $1.9141$ was used. Both figures reinforce the observations made in Section~\ref{sec:val_theory} and clearly illustrate the marginal advantage in packet transmissions that systematic network coding exhibits over non-systematic network coding for $q=2$ and high SINR values. Furthermore, Fig.~\ref{fig:opp} depicts the steep decrease in outage probability as the SINR improves. This characteristic can be attributed to the nature of the physical layer of LTE-A, which can cause sharp changes in the TB error probability when the SINR increases or reduces by $5$ dB or less beyond a particular value~\cite{sesia2011lte}.

\section{Conclusions and Future Directions}
\label{sec:conclusion}

The use of random linear network coding for the encoding of generations of source packets and the broadcast of coded packets was considered. Both the non-systematic and systematic implementations of network coding were studied and closed-form expressions for the average number of required packet transmissions, which is related to the incurred decoding delay at a receiver, were derived. Whereas previous studies focused on network coding over large fields or unconstrained packet transmission, this work looked at deadline-constrained packet transmission and finite fields of any size.

The proposed framework established that the field size has a marginal impact on the average number of required packet transmissions and the average decoding delay. Furthermore, results showed that systematic network coding can offer a small gain in delay performance over non-systematic network coding, in addition to the reported benefits of reduced computational complexity. The derived expressions and bounds can also be used to fine-tune the generation size, the field size and the maximum permissible number of transmitted packets, such that the transmitter does not waste capacity and energy in an attempt to achieve a low outage probability. 

The average decoding delay at a receiver can be considered in optimization problems that aim to minimize the energy broadcast by the transmitter or maximize the energy harvested by each receiver. For example, if the transmitter broadcasts a fixed number of coded packets per generation, a receiver could switch to energy-harvesting mode as soon as a generation is recovered. The derived expressions could be used to determine a deadline that achieves a balance between the desired outage probability and the required harvested energy. If feedback channels are available, coded packet transmissions could cease before the deadline is reached. In that case, the probability that all of the receivers will recover a generation has to be computed and used in the calculation of the average decoding delay \textit{of the system}. The delay distribution for a system consisting of one transmitter and two receivers was studied in \cite{Nistor2011}, and special cases were discussed in \cite{Nistor2011, Eryilmaz2006, Heide2009} but derivation of an exact expression remains an open problem.

\section{Acknowledgements}
\label{sec:ack}

The authors acknowledge the contribution of the COST Action IC1104. All data created during this work are openly available from the Lancaster University data archive at \url{http://dx.doi.org/10.17635/lancaster/researchdata/131}.

\appendices


\section{Equivalent expression of $f(\omega)$}
\label{apdx:equiv_PMF}

Let us first define the coefficient
\begin{equation}
C^{K+\omega-1}_m = {K+\omega-1 \choose m}\,(1-\varepsilon)^{m}\,\varepsilon^{K+\omega-1-m}\nonumber
\label{eq:coeff}
\end{equation}
and then regroup the terms of $f(\omega)$ in \eqref{eq:PMF_final} as follows
\begin{equation}
\begin{split}
f(\omega)\negsp\negsp\negsp=&\!\!\!\!\negsp\sum_{m=K}^{K+\omega-1}\!\!\!\!\bigl[(1-\varepsilon)C^{K+\omega-1}_{m-1}-(1-\varepsilon)C^{K+\omega-1}_{m}\bigl] P(m)\negsp\\
&\!+\,(1-\varepsilon)\,C^{K+\omega-1}_{K+\omega-1}\,P(K\negsp+\negsp\omega).
\end{split}
\label{eq:PMF_appdx_1}
\end{equation}
If coefficient $C^{K+\omega-1}_m$ is simultaneously added to and subtracted from $(1-\varepsilon)\,C^{K+\omega-1}_{m-1}-(1-\varepsilon)\,C^{K+\omega-1}_m$, we obtain
\begin{equation}
\begin{split}
&(1-\varepsilon)\,C^{K+\omega-1}_{m-1}- (1-\varepsilon)\,C^{K+\omega-1}_m = \\
&=\negsp (1-\varepsilon)\,C^{K+\omega-1}_{m-1}+ \varepsilon\,C^{K+\omega-1}_m- C^{K+\omega-1}_m.
\end{split}
\label{eq:expanded_coeff}
\end{equation}
Using the recursive relationship of binomial coefficients, i.e.,
\begin{equation}
{K+\omega-1 \choose m-1} + {K+\omega-1 \choose m} = {K+\omega \choose m}\nonumber
\end{equation}
we can show that
\begin{equation}
(1-\varepsilon)\,C^{K+\omega-1}_{m-1}+\varepsilon\,C^{K+\omega-1}_m=C^{K+\omega}_m.
\label{eq:coeff_reduction}
\end{equation}
If we substitute \eqref{eq:coeff_reduction} into \eqref{eq:expanded_coeff} and then into \eqref{eq:PMF_appdx_1}, and also observe that
\begin{equation}
(1-\varepsilon)\,C^{K+\omega-1}_{K+\omega-1}=C^{K+\omega}_{K+\omega}\nonumber
\end{equation}
we can rewrite $f(\omega)$ as
\begin{equation}
f(\omega)=\!\sum_{m=K}^{K+\omega}\!\!C^{K+\omega}_{m}P(m)-\!\!\sum_{m=K}^{K+\omega-1}\!\!C^{K+\omega-1}_{m}P(m)
\label{eq:PMF_appdx_2}
\end{equation}
where the first and second sums in \eqref{eq:PMF_appdx_2} correspond to $F(\omega)$ and $F(\omega\negsp-\negsp1)$, respectively.


\section{Proof of Corollary \ref{pro:avg_buf_delay_unconstrained}}
\label{apdx:proof_upper_bound}

Based on Definition \ref{def:avg_delay}, the average decoding delay is given by $\bar{d}=K+\lim_{\Omega\rightarrow\infty}\bar{\omega}$. Taking into account that $F(\nu)$ can be expanded into $f(0)+\ldots+f(\nu)$ for $\nu=0,\ldots,\Omega\!-\!1$, the average overhead $\bar{\omega}$ given in \eqref{eq:delay_NC_fundamental} can be written as 
\begin{equation}
\begin{split}
\bar{\omega}&=\Omega-\sum\limits_{\ell=0}^{\Omega-1}f\!\left(\ell\right)\,\left(\Omega-\ell\right)\\
&=\Omega-\Omega\sum\limits_{\ell=0}^{\Omega-1}f\!\left(\ell\right)+\sum\limits_{\ell=0}^{\Omega-1}\ell\;f\!\left(\ell\right).
\end{split}
\label{eq:avg_delay_decomposed}
\end{equation}
For $\Omega\rightarrow\infty$, the second sum in the last line of \eqref{eq:avg_delay_decomposed} represents the probability that the $K$ packets will be recovered after an infinite number of time steps have elapsed or, equivalently, an infinite number of packet transmissions have occurred. This probability is equal to 1 for non-zero erasure probability values. Consequently, the first and second terms in \eqref{eq:avg_delay_decomposed} cancel each other out and the limit of $\bar{\omega}$ as $\Omega\rightarrow\infty$ reduces to
\begin{equation}
\lim_{\Omega\rightarrow\infty}\bar{\omega}=\lim_{\Omega\rightarrow\infty}\left[\,\sum\limits_{\ell=1}^{\Omega-1}\ell\;f\!\left(\ell\right)\,\right].
\label{eq:basic_limit}
\end{equation}
Note that the starting value of $\ell$ has been set to 1 because the first term of the sum in \eqref{eq:basic_limit} is zero for $\ell=0$.

Let us define $\Delta(\nu)=P(K+\nu)-P(K+\nu-1)$ for convenience, where $\Delta(0)=P(K)$. Consequently, \eqref{eq:PMF_final} can be rewritten as
\begin{equation}
f\!\left(\ell\right)=\sum_{r=0}^{\ell}{K+\ell-1 \choose \ell-r}\varepsilon^{\ell-r}\left(1-\varepsilon\right)^{K+r}\Delta(r).
\label{eq:PMF_shortdef}
\end{equation}
If we write $\ell$ as $(\ell-r)+r$ and invoke \eqref{eq:PMF_shortdef}, the sum in \eqref{eq:basic_limit} can be decomposed into two sums, denoted by $\Sigma_1$ and $\Sigma_2$, as follows
\begin{equation}
\begin{split}
&\sum\limits_{\ell=1}^{\Omega-1}\ell\,f\!\left(\ell\right)=\\ 
&\underbrace{\sum\limits_{\ell=1}^{\Omega-1}\sum\limits_{r=0}^{\ell-1}\left(\ell-r\right)\!{K+\ell-1 \choose \ell-r}(1-\varepsilon)^{K+r}\varepsilon^{\ell-r}\,\Delta(r)}_{\!\!\displaystyle\phantom{\bar{\Sigma}}\Sigma_1}+\\ 
&+\underbrace{\sum\limits_{\ell=1}^{\Omega-1}\sum\limits_{r=1}^{\ell}r{K+\ell-1 \choose \ell-r}(1-\varepsilon)^{K+r}\varepsilon^{\ell-r}\,\Delta(r)}_{\!\!\displaystyle\phantom{\bar{\Sigma}}\Sigma_2}.
\end{split}
\label{eq:sum_decomposition}
\end{equation}
Note that, in order to discard zero terms, the summation index $r$ stops at $\ell-1$ in $\Sigma_1$ and starts from $1$ in $\Sigma_2$.

The sum $\Sigma_1$ collapses to
\begin{equation}
\Sigma_1=\sum\limits_{\ell=1}^{\Omega-1}\varepsilon\,(K+\ell-1)\,f\!\left(\ell\!-\!1\right)\nonumber
\label{eq:sum_1v1} 
\end{equation}
if we observe that components of $\Sigma_1$ in \eqref{eq:sum_decomposition}  can be rearranged to form $f\!\left(\ell\!-\!1\right)$, as defined in \eqref{eq:PMF_shortdef}. Changing the summation index from $\ell$ to $j=\ell-1$ gives
\begin{equation}
\Sigma_1=\varepsilon\,K\sum\limits_{j=0}^{\Omega-2}f\!\left(j\right)+\varepsilon\sum\limits_{j=0}^{\Omega-2}j\,f\!\left(j\right).
\label{eq:sum_1v2}
\end{equation}

If we return our attention to \eqref{eq:sum_decomposition}, we observe that the order of summation in $\Sigma_2$ can be interchanged and the limits of the summation indices $\ell$ and $r$ can be updated accordingly, resulting in 
\begin{equation}
\Sigma_2\negsp=\!\sum\limits_{r=1}^{\Omega-1}\negsp r\,\negsp\Delta(r)\sum\limits_{\ell=r}^{\Omega-1}\negsp{K+\ell-1 \choose \ell-r}(1-\varepsilon)^{K+r}\varepsilon^{\ell-r}.
\label{eq:sum_2v1}
\end{equation}
Careful inspection of the expression in the inner sum of \eqref{eq:sum_2v1} reveals that it corresponds to a PMF. More specifically, let us consider a source transmitting $\kappa$ uncoded packets to a destination in $\kappa$ time steps over a point-to-point channel characterized by erasure probability $\varepsilon$. The destination uses a feedback channel to notify the source of erased packets. Subsequently, the source dedicates $\delta$ additional time steps for packet retransmissions. The probability that the destination will recover the $\kappa$ packets in \textit{exactly} $\kappa+\delta$ time steps is
\begin{equation}
\theta(\delta, \kappa)={\kappa+\delta-1 \choose \kappa-1}\;(1-\varepsilon)^{\kappa}\;\varepsilon^{\delta}.
\label{eq:PMF_uncoded_ptop}
\end{equation}
This is because the $\kappa$-th packet will be recovered in time step $\kappa+\delta$ and the remaining $\kappa-1$ packets have been retrieved in the previous $\kappa+\delta-1$ time steps. Invoking the PMF in \eqref{eq:PMF_uncoded_ptop} and changing the summation index of the inner sum in \eqref{eq:sum_2v1} from $\ell$ to $i=\ell-r$, we write $\Sigma_2$ in the following form
\begin{equation}
\Sigma_2=\sum\limits_{r=1}^{\Omega-1}\! r\,\negsp\Delta(r)\!\!\sum\limits_{i=0}^{\Omega-1-r}
\!\theta\!\left(i,\,K+r\right).\nonumber
\label{eq:sum_2v2}
\end{equation}
The inner sum of PMFs is bounded from above by 1, while $\Delta(r)$ can be expanded to $P(K+r)-P(K+r-1)$ giving
\begin{equation}
\begin{split}
\Sigma_2&<\sum\limits_{r=1}^{\Omega-1} r\,\bigl[P(K+r)-P(K+r-1)\bigr]\\
&=(\Omega-1)\,P(K+\Omega-1)-\sum_{r=0}^{\Omega-2}P(K+r).
\end{split}
\label{eq:sum_2v3}
\end{equation}
A lower bound on the probability $P(K+r)$, given in \eqref{eq:full_rank}, can be computed as follows
\begin{equation}
\resizebox{.15\textwidth}{!}
{$\displaystyle\prod_{\lambda=0}^{K-1}\!\!\left(1\negsp-q^{-K-r+\lambda}\right)\negsp$}
\geq
\resizebox{.185\textwidth}{!}
{$\displaystyle 1-q^{-r-1}\left(\frac{1-q^{-K}}{1-q^{-1}}\right).$}
\nonumber
\label{eq:rank_prob_bound}
\end{equation}
Therefore, the sum in the second line of \eqref{eq:sum_2v3} can also be bounded:
\begin{equation}
\begin{split}
\resizebox{.127\textwidth}{!}
{$\displaystyle\sum_{r=0}^{\Omega-2}\!P(K+r)$}
&\geq 
\resizebox{.242\textwidth}{!}
{$\displaystyle(\Omega-1)-\left(\frac{1-q^{-K}}{1-q^{-1}}\right)\!\sum_{r=0}^{\Omega-2}q^{-r-1}$}\\
&=
\resizebox{.283\textwidth}{!}
{$(\Omega-1)-\displaystyle\frac{q^{-1}\left(1-q^{-K}\right)\!\left(1-q^{-\Omega+1}\right)}{\left(1-q^{-1}\right)^2}.$}
\end{split}
\label{eq:sum_rank_prob_bound}
\end{equation}
Recall that, according to \eqref{eq:sum_decomposition}, the sum $\sum_{\ell=1}^{\Omega-1}\ell\,f\!\left(\ell\right)$ is equal to $\Sigma_1+\Sigma_2$, where $\Sigma_1$ was computed in \eqref{eq:sum_1v2} and an upper bound on $\Sigma_2$ can be obtained if we substitute \eqref{eq:sum_rank_prob_bound} into \eqref{eq:sum_2v3}. Taking the limit of \eqref{eq:sum_decomposition} as $\Omega\rightarrow\infty$, we find that the average overhead is bounded from above by
\begin{equation}
\lim_{\Omega\rightarrow\infty}\bar{\omega}<\left(\frac{1}{1-\varepsilon}\right)\left[\varepsilon K+\frac{q^{-1}(1-q^{-K})}{(1-q^{-1})^2}\right].
\label{eq:upper_bound}
\end{equation}
Adding $K$ to the right-hand side of \eqref{eq:upper_bound} gives the lower bound on the average decoding delay $\bar{d}$.

\bibliographystyle{IEEEtran}
\bibliography{IEEEabrv,IEEE_TVT_ICandAT_ArXiv_Ref}

\end{document}